# Enhanced graphitic domains of unreduced graphene oxide and the interplay of hydration behaviour and catalytic activity

*Tobias Foller[1], Rahman Daiyan[2], Xiaoheng Jin[1], Joshua Leverett[2], Hangyel Kim[3], Richard Webster[4], Jeaniffer E. Yap[1], Xinyue Wen[1], Aditya Rawal[4], K. Kanishka H. DeSilva[5], Masamichi Yoshimura[5], Heriberto Bustamante[6], Shery L.Y. Chang[1,4], Priyank Kumar[2], Yi You[1,7], Gwan-Hyoung Lee[3], Rose Amal[2] and Rakesh Joshi[1],\**

[1]School of Materials Science and Engineering, University of New South Wales, Sydney, NSW 2052, Australia

[2]Particles and Catalysis Research Laboratory and School of Chemical Engineering, University of New South Wales, Sydney, NSW, 2052, Australia

[3]Department of Materials Science and Engineering Research Institute of Advanced Materials (RIAM), Seoul National University, Seoul 08826, Korea

[4] Electron Microscopy Unit, Mark Wainwright Analytical Centre, University of New South Wales, Sydney, NSW, 2052, Australia

[5]Surface Science Laboratory, Toyota Technological Institute, Nagoya 468-8511, Japan

[6]Sydney Water, Parramatta, New South Wales 2125, Australia

[7]School of Physics and Astronomy, University of Manchester, Manchester M13 9PL, UK

\*email: r.joshi@unsw.edu.au

**Abstract**

*Previous studies indicate that the properties of graphene oxide (GO) can be significantly improved by enhancing its graphitic domain size through thermal diffusion and clustering of functional groups. Remarkably, this transition takes place below the decomposition temperature of the functional groups and thus allows fine tuning of graphitic domains without compromising with the functionality of GO. By studying the transformation of GO under mild thermal treatment, we directly observe this size enhancement of graphitic domains from originally ≤ 40 nm$^2$ to > 200 nm$^2$ through an extensive transmission electron microscopy (TEM) study. Additionally, we confirm the integrity of the functional groups during this process by a comprehensive chemical analysis. A closer look into the process confirms the theoretical predicted relevance for the room temperature stability of GO. We further investigate the influence of enlarged graphitic domains on the hydration behaviour of GO and catalytic performance of single atom catalysts supported by GO.*

Graphene oxide (GO) is a versatile material with a broad spectrum of possible applications such as water purification, gas separation, energy storage, and energy harvesting [1–5]. Mild thermal treatment of GO revealed itself as a powerful tool to further fine tune and improve various properties of GO by enhancing the graphitic domain size without reducing GO[3,6–10]. These studies explain the improvements of GO by a thermally driven agglomeration of the functional groups on the GO basal plane. Mild temperatures (around 80 °C) allow the functional groups to diffuse and cluster. As a consequence a pronounced ordering of the sp$^2$/sp$^3$ phase (graphitic/functionalized domains) is achieved, increasing the size of graphitic domains without compromising with the oxygen content [3,6–8,11]. Moreover, various theoretical works support facile diffusion and energetically favourable clustering of oxygen functionalities and predict its key role for the room temperature stability of GO [3,6–8,11–15]. Despite its importance from a fundamental science as well as potential application point of view, there is still a lack of a direct experimental observation showing the clustering of oxygen containing functional groups in thermally treated GO. In this study, we observe the enhancement of the graphitic domain size and the conservation of functional groups in transmission electron microscopy TEM. We further examine the integrity of the functional groups with solid state nuclear magnetic resonance (SSNMR), x-ray photoelectron spectroscopy (XPS) and Fourier-transform infrared spectroscopy (FTIR).

Tailoring the graphitic domains of GO without largely changing the amount of oxygen containing functionalities, allows to investigate the influence of graphitic domain size on the properties of GO. It is for example predicted that graphitic domains represent patches along which the water transport is frictionless[16]. Thus, interconnected graphitic domains should facilitate the water transport along laminated GO sheets, as used in water purification applications. Here, we experimentally and theoretically analyse the water transport in GO with enlarged graphitic domains and find that the water transport is gradually decreased despite a larger mobility of water molecules along the GO planes as demonstrated in our ab-initio molecular dynamics (AIMD) simulations.

The integrity of the oxygen functional groups and control over the hydration behaviour now allows to further investigate the properties of GO. In recent years, laminated 2D materials as a catalyst itself or as support for single atom catalysts has emerged as a rapidly growing research field[17–20]. Here, we use an excellent performing single Fe ion oxygen evolution reaction (OER) catalyst supported by GO laminates to investigate the influence of enlarged graphitic domains and hydration behaviour on the catalytic performance. By that, we reveal the critical role of water transport in the activity of catalysts based on laminar 2D materials.

**Results and discussion**

**Observation of thermally enhanced graphitic domains**

To examine the crystal structure and chemical composition of thermally treated GO, patches of GO covered TEM grids, GO powder, and GO films were annealed at 80 °C for several days in vacuum and air. The annealing temperature of 80 °C was chosen as previous studies indicate that this temperature is optimal for the agglomeration process[3,6–8,11]. It is below the decomposition temperature of functional groups and leads to most prominent changes in the properties of GO [6]. Annealing in vacuum and air is compared due to the fact that previous studies report different results about the conservation of the oxygen content in mild thermal annealing[6,11]. However, in one of the studies the annealing is performed in air, in the other in vacuum environment. Therefore, in this study both environments are compared to investigate whether the discrepancy in the literature originate from this parameter.

It is possible to classify TEM images of GO monolayer by high contrast regions as disordered areas containing oxygen functionalities and low contrast regions as ordered areas representing graphitic domains[21,22]. Therefore, by TEM imaging the state and distribution of the $sp^2/sp^3$ (graphitic/functionalized) phase can be directly observed. The objective of this study is now to examine whether the mild thermal treatment leads to an increase of graphitic domain size while preserving the amount of functional groups, i.e. enhancing the size of crystalline domains without reducing the GO. For that, a time series of annealed GO samples is investigated and analysed with a statistical approach to quantify the agglomeration process. Fig. 1 A/B exemplarily show a time series of TEM images of untreated GO monolayer (Day 0) and heat-treated GO (80 °C, Day 1 –

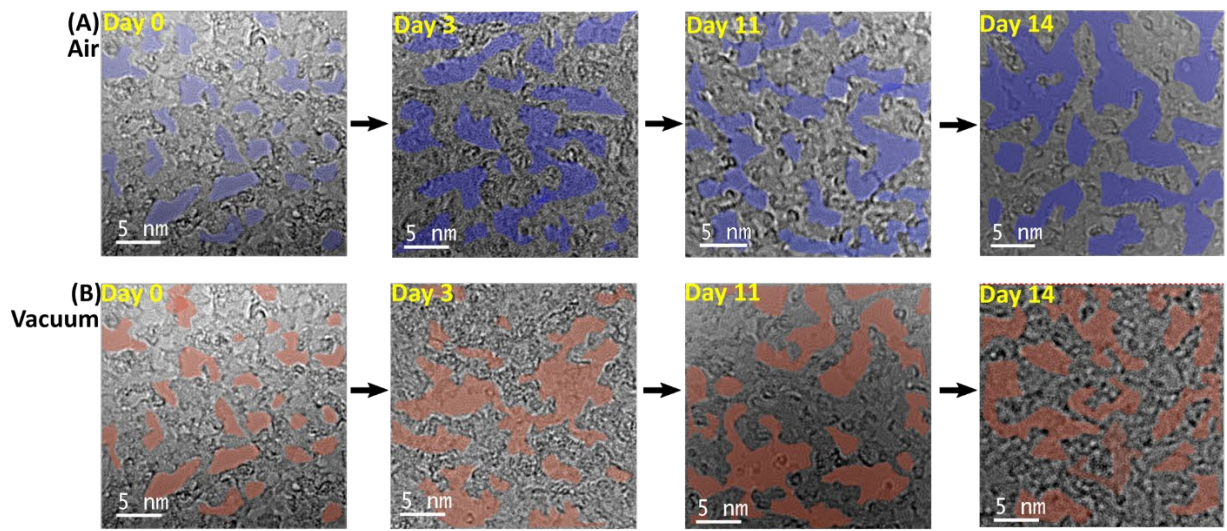
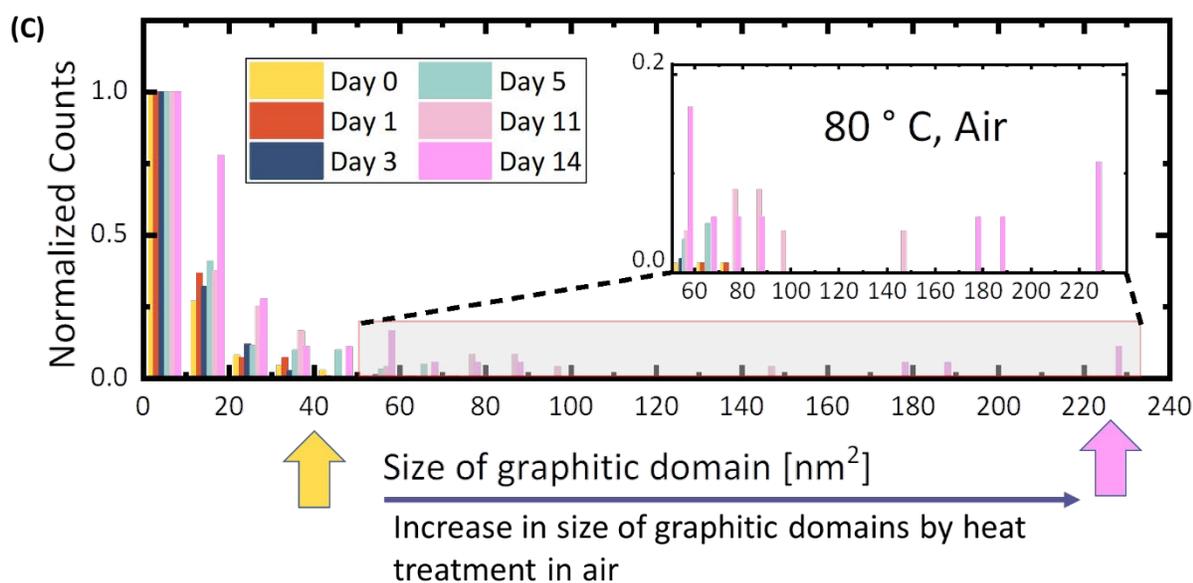
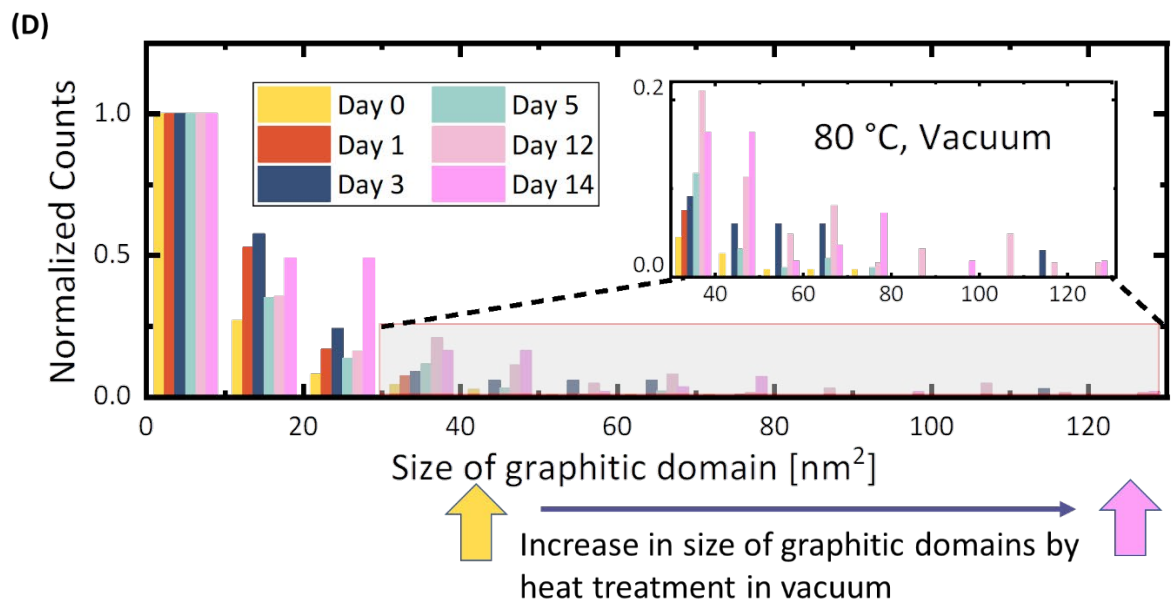

**Figure 1 Observation of thermally enhanced graphitic domains** (A/B) Time course of GO after heat treatment in air/vacuum. Graphitic domains are marked in light blue/red for treatment in air/vacuum C/D) Histogram of the size distributions of graphitic domains for increasing days of heat treatment in air/vacuum. Day 0 samples represent the same untreated GO. To illustrate the enhancement of graphitic domains, the maximum size of graphitic domains in untreated GO is marked with a yellow arrow (~ 40 nm$^2$) and the maximum size of 14 days heat treated GO is marked with a pink arrow. (~ 230 nm$^2$ in air and ~ 120 nm$^2$ in vacuum)

Day 14 in air/vacuum). For clarity, well-ordered areas (graphitic domains) are false coloured with light blue/red while disordered areas (functionalized areas) remain uncoloured. Since this distinction is only possible for monolayer GO, only monolayer regions were used for the analysis identified by selected area diffraction (Fig. S1). Consistent with previous studies [21], the untreated GO show graphitic domains with a size of mainly 1-10 nm$^2$ and up to 40 nm$^2$. However, the GO samples annealed for 14 days in air and vacuum show fewer but larger graphitic domains with sizes up to 230 nm$^2$ and 120 nm$^2$. Therefore, the heat-treated GO shows a clear increase in size of graphitic domains in both environments. Exemplarilly, GO anealed for 14 days in air was investigated with abberration corrected TEM (Fig. S2). The TEM images clearly show large graphitic areas of six rings surrounded by functionalized and defective areas containing four and five rings. This supports our methodeology of observing the state and distribution of the sp2/sp3 phase in annealed GO samples.

For a representative view of the changes during heat treatment, a statistical approach averaging over 6-12 images per heat treatment day was undertaken. Analysing the development of the number of separate graphitic domains, their size distribution as well as the percentage coverage of graphitic area per image allows to quantify a potential ordering process upon mild thermal heat treatment (see SI – 1.3, Fig. S3 for analysis procedure of size, number, and coverage of graphitic domains). If an ordering process occurs without significantly reducing the number of functional groups, it should manifest in the following observables. The number of graphitic domains should decrease while their size should increase both as a consequence of graphitic domains joining over the course of heat treatment. Meanwhile the percentage coverage of graphitic area should remain constant since the area of functionalized graphene does not decrease. Fig. 1C-D and S4A, S4B, summarize these findings for heat treatment in air and vacuum. In both environments, the number of graphitic domains per image (27 nm x27 nm) decreases exponentially with time (Fig. S4A), indicating that graphitic domains join during the heat treatment process. Simultaneously, the average percentage coverage with graphitic area remains constant within the measurement capabilities (Fig. S4B). Consequently, the area of functionalized graphene also stays constant within that limit too. The histogram in Fig. 1C of the size distribution shows that the size of the graphitic domains subsequently increases with time, giving rise to graphitic domains up to the range of 150-250 $nm^2$ after 14 days of heat treatment in air and 80-130 $nm^2$ after 14 days heat treatment in vacuum (Fig. 1D) again showing the joining of graphitic domains. As mentioned above, such large graphitic domains are not observable in untreated GO (< 40 $nm^2$). Comparing the heat treatment in air and vacuum reveals that the number of graphitic domains decrease faster but saturates quicker in a higher amount of smaller graphitic domains in vacuum than in air (see Fig. 1A-D and S4A-B).

The observed enhancement of the size of the graphitic domains could also originate from a removal of adsorbates such as hydrocarbon and physiosorbed oxygen as well as from the decomposition of functional groups by either the mild thermal treatment or electron exposure while recording the TEM images [21,23]. The fact that the percentage of graphitic area do not undergo drastic changes under mild thermal treatment (Fig. S4B) and electron exposure (Video S1), suggests that the changes do not mainly originate from the decomposition of functional groups or the removal of adsorbates as discussed in detail in SI – 1.4. However, more subtle changes to the composition and number of functional groups do not lie within the sensitivity of this measurement technique.

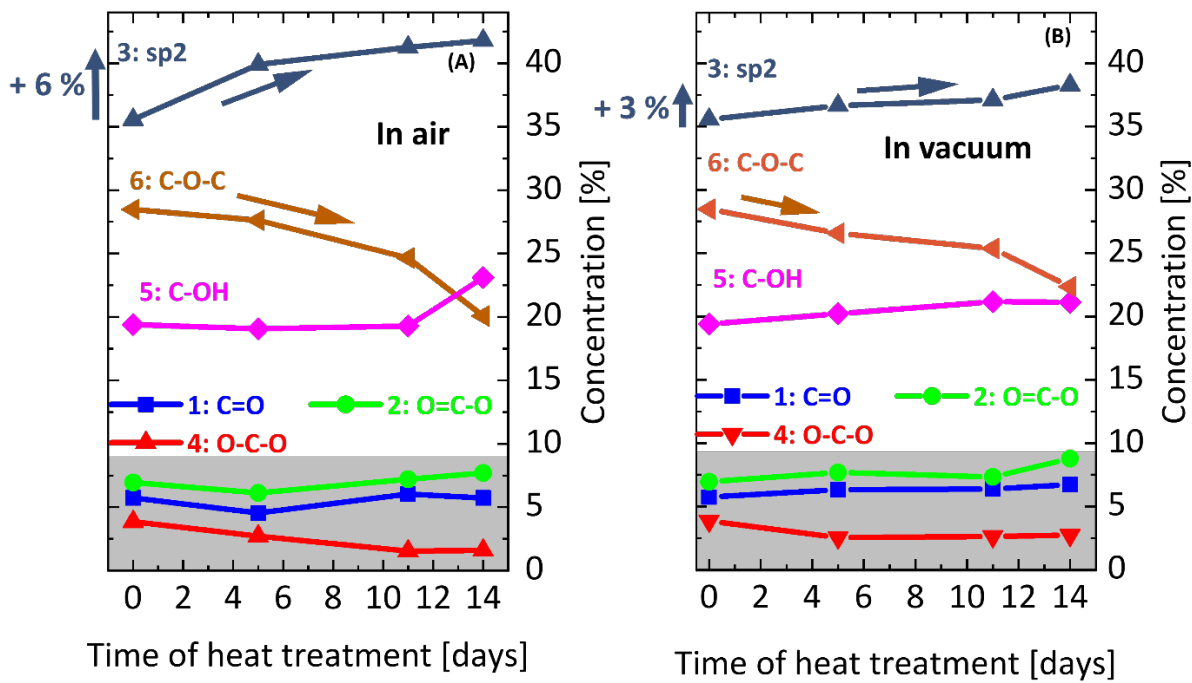

**Figure 2 SSNMR study of annealed GO (A/B)** Concentration of functional groups for GO annealed in **air/vacuum** from SSNMR analysis of annealed GO. Peaks 1-6 are assigned following previous studies[24,25]. A slight increase of 6 % and 3 % of sp$^2$ carbon in air and vacuum are highlighted in the graph. Error bars were depicted from signal to noise ratio. The peaks in the grey background have a signal to noise ratio which does not allow quantifiable conclusions.

Therefore, solid state nuclear magnetic resonance (SSNMR) measurements over the course of the heat treatment on GO powder were performed. Following previous studies [24,25], the three most prominent peaks around 60, 70 and 130 ppm are assigned to epoxy (C-O-C), hydroxyl (C-OH) and sp$^2$ carbon. The three additional, low intensity peaks are assigned to lactol (~100 ppm), carboxylic acid (~167 ppm) and ketone (~190 ppm). As shown in Fig. 2A an increase of sp$^2$ carbon by 6 % over 14 days of heat treatment in air can be observed which is dominantly in the first 5 days and becomes less with ongoing heat treatment. Simultaneously, the relative intensity of the epoxy (C-O-C) peak (60 ppm) decreases by 7 % and hydroxyl (C-OH, 70 ppm) increases by 3 %. Heat treatment of GO in vacuum causes changes in composition of functional groups that are less pronounced than in air (see Fig. 2B). The sp2 carbon increases by 3 %, while epoxy groups decrease by 6 % and hydroxyl groups increase by 2 %.

According to the Lerf-Klinowski-model for GO from Hummers' method, epoxy and hydroxyl groups are prominent on the basal plane, while carboxylic acid, ketone and lactol are mostly present on the edges[24–27]. Thus, to investigate the behaviour of functional groups on the basal plane upon heat treatment the latter groups are neglected in the further discussion. Additionally, Fourier-transform infrared spectroscopy (FTIR) and X-ray photoelectron spectroscopy (XPS) studies of the heat-treated samples were performed which confirm marginal changes in the C/O ration and composition of functional groups upon heat treatment in air and vacuum (SI – 2).

We can conclude that all our analytical techniques, consistently show that the loss of chemically bond oxygen functionalities is marginal compared to the observed large increase in graphitic domain sizes in TEM. Thus, we mainly attribute the enhancement of graphitic domains to the diffusion and agglomeration of functional groups and not to a reduction of GO. As discussed in detail in SI-3, our combined observation of increasing graphitic domain size and decreasing decomposition rate of functional groups with respect to heat treatment time, supports the theoretical prediction that the agglomeration process is beneficial for the room-temperature stability of GO [15]. The comparison between heat treatment in air and vacuum also supports our findings since the results are similar and thus reproduced. Interestingly, the results differ in detail. According to our TEM results, the enhancement of graphitic domains takes place faster but is less pronounced in vacuum than in air. As shown with our SSNMR analysis, the increase in sp$^2$ carbon is less pronounced in vacuum with 3 % compared to 6 % in air, indicating that the decomposition of functional groups is damped in vacuum compared to air. This might contribute to the fact that the observed graphitic domains are larger in air than in vacuum after 14 days of heat treatment. However, it does not explain why the agglomeration in vacuum is slightly faster than in air. Therefore, our observations show, that the diffusion, agglomeration and decomposition process in air and vacuum are different. A detail, that did not get attention in the literature yet and thus need further experimental and theoretical investigations.

**Influence of enlarged graphitic domains on hydration behaviour of GO**

As mentioned in the introduction, enhanced graphitic domains should facilitate the water transport along stacked GO laminates due to frictionless water flux. Therefore, the hydration behaviour of heat-treated GO was investigated. In order to investigate the hydration behaviour, a study of water permeating GO and its interlayer space was conducted and supported by ab initio molecular dynamics (AIMD) simulations (Fig. 3 A-E). Fig. 3A shows the humidity dependent water loss of a container sealed with a GO membrane. Original GO, 7- and 14-days heat treated membranes show an exponential increase of water loss with humidity. This is typical for GO membranes. With higher humidity more water enters the spaces between the sheets. After filling up the voids in graphitic areas the membrane starts to swell [16,28]. The water loss clearly decreases with increasing time of heat treatment, especially in higher humidity. In 100 % RH the weight loss of the containers is decreased by a factor ~2 and ~4 for 7 and 14 days of heat treatment in air. Measured in ambient air, the interlayer space after the heat treatment decreases linearly from 0.75-0.65 nm (Fig. S7). Similarly, the wet interlayer space of GO is also decreased with increasing time of heat treatment (Fig. 3B). This supressed swelling upon annealing of GO is discussed in detail in the supporting information (SI – 4.1).

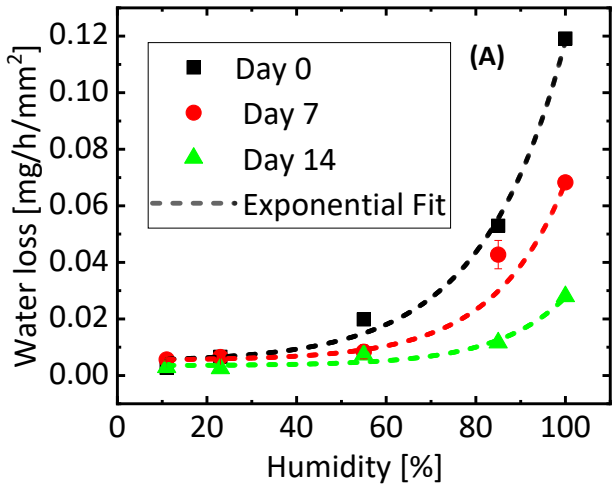
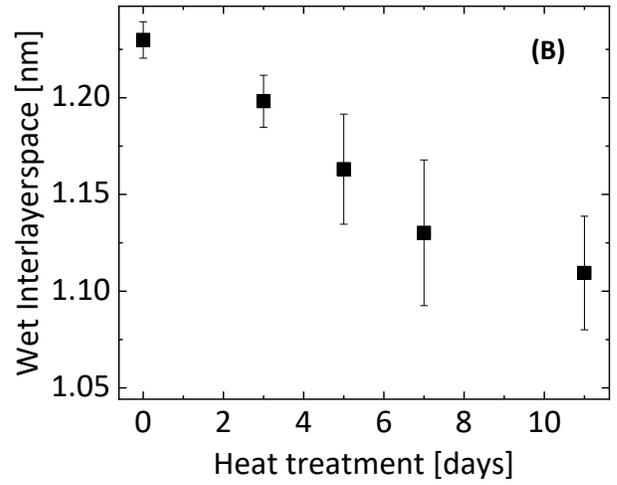
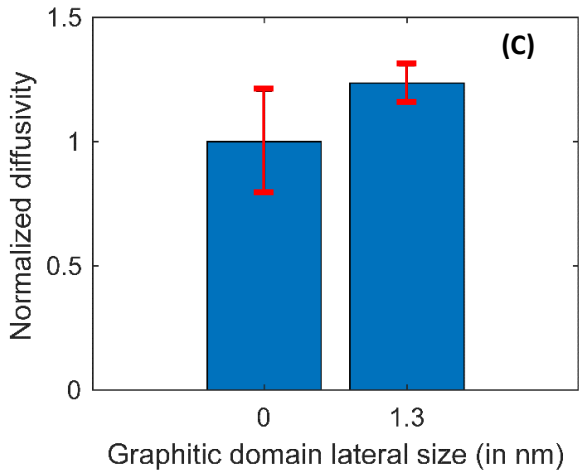
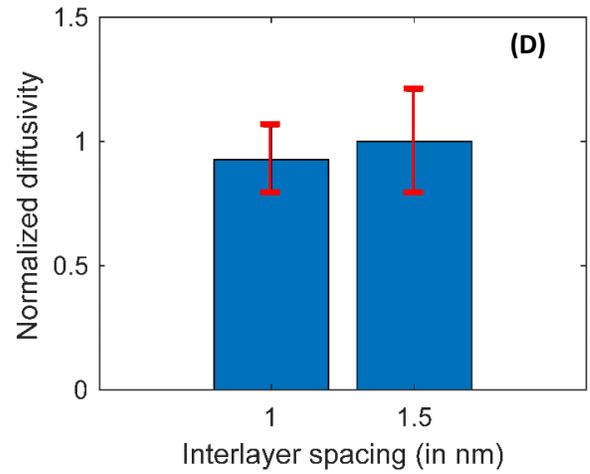
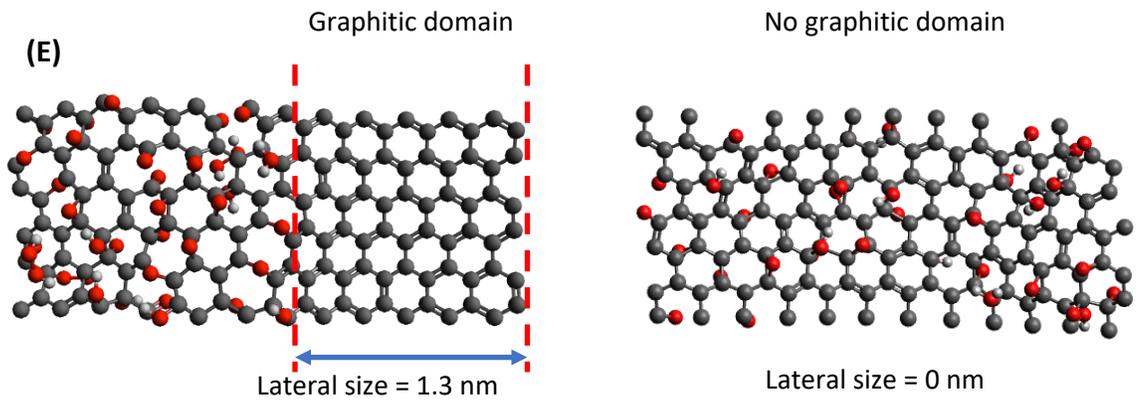

**Figure 3 Hydration behavior of heat-treated GO.** A) Water loss of a container sealed with GO membranes heat-treated for 0, 7 and 14 days. The humidity inside the container was controlled with saturated salt solutions. B) Wet interlayer space of heat treated GO. C, D) Diffusivity values obtained for water molecules placed within the interlayer space. The effect of both graphitic domain size and interlayer spacing between GO structures is shown. The diffusivity values are normalized with respect to the mean value obtained for the 0 nm graphitic domain case C) and the mean value obtained for the 1 nm interlayer spacing case D). The error bars represent the standard error of the mean calculated for three cases. E) Example structural models of GO utilized in our AIMD simulations, highlighting structures with and without the presence of graphitic domains. The lateral domain sizes of 1.3 nm and 0 nm cases are indicated.

To understand the dual effect of interlayer spacing (degree of swelling) and the enhancement of graphitic domains in GO on water flux, we carried out AIMD simulations of water-GO systems (Fig. 3 C-E). We employed the diffusivity as a proxy for the water flux. As such, a higher (lower) value of the MSD indicates a higher (lower) water flux. First, we compared the diffusivity values of water molecules that move in between GO sheets with varying size of graphitic domains (graphitic domain lateral sizes of 0 nm, i.e., no graphitic domain and 1.3 nm, see Fig. 3E), but with a fixed interlayer spacing (1.5 nm). Our computational results show that larger graphitic channels lead to a higher diffusivity value of water molecules (Fig. 3C), thus implying faster water transport along such structural motifs. Second, we compared the diffusivity values of water molecules that move in between GO sheets with varying interlayer spacings (1.0 and 1.5 nm spacing between the carbon planes) for a fixed graphitic domain size (0 nm). These results, shown in Fig. 3D, reveal that GO sheets with smaller interlayer spacings (suppressed swelling) lead to relatively lower diffusivity values, thus indicating slower water transport in GO structures with suppressed swelling. Given that our experimental results showed an overall decrease in the water flux with the time of heat treatment, we can conclude that suppressed swelling or smaller interlayer spacings play a more dominant role in controlling the water transport. Thus, the supressed swelling of the heat-treated GO membranes gradually hinders the water transport inside the GO capillaries.

**Electrochemical analysis of enhanced graphitic domains and annealed GO-SAC**

Larger graphitic domains might improve the HER and OER activity of GO due to an increase in electrode conductivity [6], resulting in faster electron transfer from the electrode to the reactants. To investigate that, we performed electrochemical experiments exploring the HER and OER of GO in alkaline environment. The performance of our catalysts is quantified by assigning an overpotential value η for which the current density $j > 10$ mA cm$^{-2}$. As shown in Fig. 4A for original GO, $j < 10$ mA cm$^{-2}$ for HER and OER. Note that the low j is further confirming the absence of any catalytically active trace metal impurities within original GO. After heat treatment, the GO starts to show catalytic activity for HER (η = 900 mV) confirming that the enlarged graphitic domains are beneficial for the HER activity.

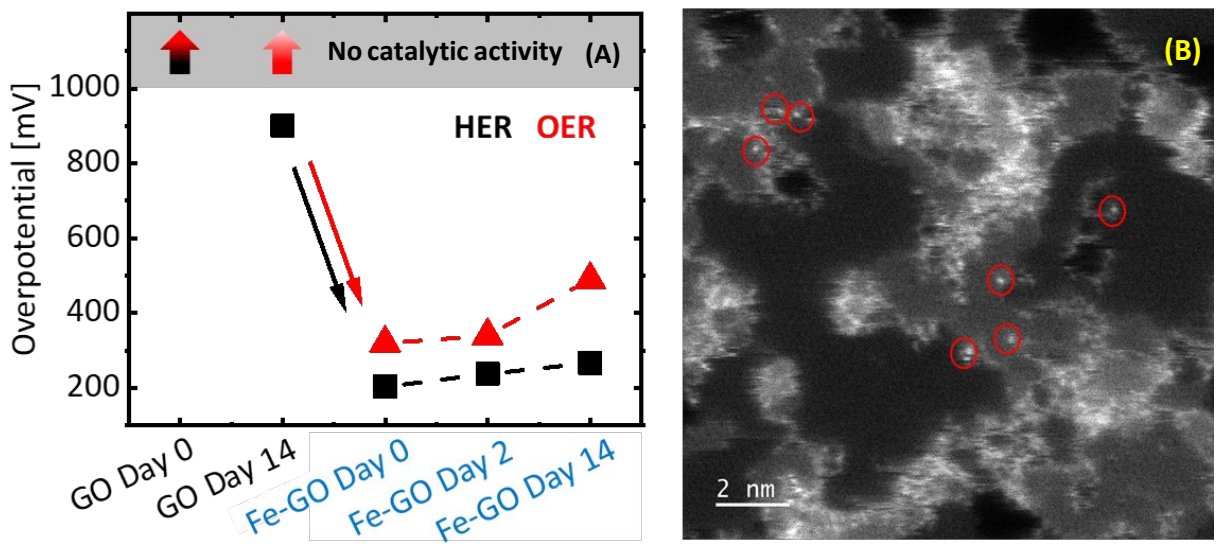

**Figure 4 Electrocatalytic activity and characterization of annealed Fe-GO** A) Alkaline HER and OER overpotential values (j > 10 mA) for GO and Fe-GO after 2 and 14 days of heat treatment. Up facing arrows in the grey shaded area indicate no catalytic activity (j < 10 mA) for HER (black) and OER (red). B) ADF-STEM image of Fe-GO. Bright spots are attributed to Fe-ions and marked with red rings.

It is now tempting to use GO as a substrate for SAC [29] and investigate the influence of enlarged graphitic domains on such a system. Based on the procedure in our previous study we intercalated Fe ions into untreated GO [5] resulting in a mix matrix that is stable when immersed in water (SI – 5.1). Annular dark field – scanning transmission electron microscopy (ADF-STEM) images of Fe-GO confirm that the Fe is dispersed as single ions within the GO network (Fig. 4B, SI – 5.2). Notably, the Fe ions are present within the functionalized area of the GO structure and rather than on the graphitic domain, suggesting a possible coordination with oxygen or defects. For as synthesized Fe-GO, $j$ is dramatically increased for alkaline HER and OER, e.i $\eta$ is ~204 mV and 320 mV. For OER this is comparable state-of-the art $IrO_2$ ($\eta$ = 270 mV) [30] and non-precious metal catalysts ($\eta$ = 300-350 mV)[31]. After initially showing OER activity of $\eta$ = 320 mV, it drastically declines for samples with 2 and 14 days of heat treatment. Similarly, subsequent annealing of Fe-GO leads to a decline in HER activity. This is surprising as it is contrary to the improvement of HER activity after heat treatment of GO (Fig. 4A). As the annealing of Fe-GO might lead to agglomeration or other changes in the Fe-ions, control samples for which the Fe ions were added after annealing the GO solution in an 80 °C oil bath for two days were investigated. The decline in HER and OER activity is similar to the decline of Fe-GO that was annealed after mixing the components. Therefore, the reason for the decline must have another reason.

The reduced water transport that was demonstrated by our study of the hydration behaviour of annealed GO might explain the decrease in catalytic activity. Fig. **5** schematically shows the mass transport of the reactants to/ from (dashed blue/orange lines) the catalytic active Fe-sites (red dots) along the GO sheets (black lines). The interlayer space $d_1$ (~1.1 nm) in annealed GO is smaller compared to the interlayer space in original GO $d_2$ (~1.3 nm). Smaller d leads to less physical space for water and consequently reduces the amount of water within the laminar network. As shown by our AMID simulations, the water transport is additionally hindered by damped diffusivity of water. Especially the transport of $OH^-$ and $H_3O^+$ rely on the hydrogen network of the water within GO laminates [33]. Hence, a slower water transport and less water within the laminates might hinder the effective transport of reactants to/from the active sites deeper within the laminar structure. This might decrease the effectiveness of the OER/HER reaction and explain the observed decrease in catalytic activity of annealed Fe-GO. Future experiments, that keep the water transport high after heat treatment might give more insight into the validation of this hypothesis. This may be achieved by intercalating additional spacers into GO that increase the interlayer space or by creating defects that facilitate the water transport.

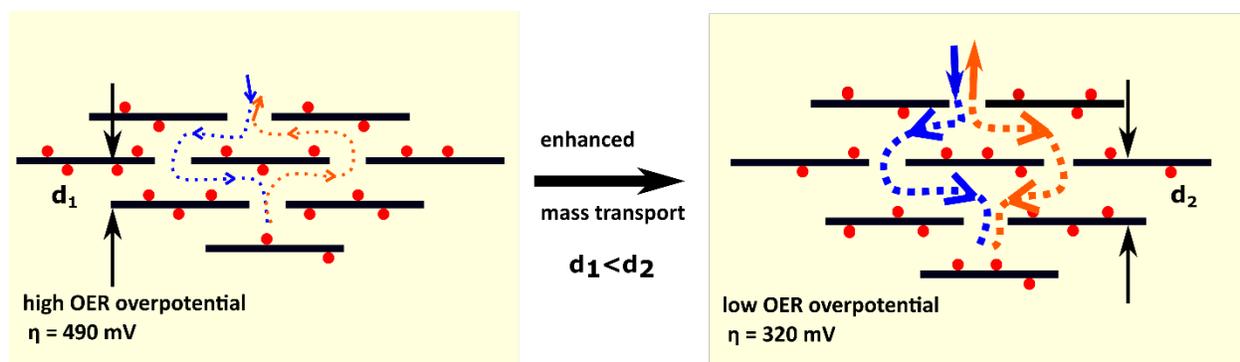

**Figure 5 Role of mass transport for the catalytic activity of GO-SAC:** Enhanced mass transport of the reactants to/from (dashed blue/orange lines) the catalytic Fe-sites (red dots) along the GO sheets (black lines) potentially increases the catalytic activity of the GO-SAC. The high/low OER overpotential of 14-days annealed/untreated (490 mV/ 320 mV) Fe-GO is correlated with low/high water transport observed for 14-days annealed/original GO. The decrease in water flux is due to a decrease in wet interlayer space d from $d_2$=1.3 nm to $d_1$=1.1 nm

**Conclusions**

The first direct observation of enlarged graphitic domains by heat treatment with temperatures under the decomposition temperature of functional groups is presented. This supports various theoretical predictions[3,6,8,12–15] with experimental data and helps to understand and control an easily scalable approach to enhance the properties of GO without compromising with the benifites of a highly functionalized material [6–11]. However, a slight decomposition of functional groups in air atmosphere is observed that declines with increased ordering upon mild thermal heating, confirming that the ordering process of functional groups contributes to the stability of GO at moderate temperatures in air [12,15]. The comparison of heat treatment in air and vacuum shows that the decomposition of functional groups is largely supressed in vacuum. This explains the seemingly contradictory results in the literature that reported both the conservation[6] and decrease[11] of functional groups upon mild thermal treatment. However, the influence of the environment was not taken into account in both of these studies.

Our investigation of the hydration behaviour of GO with enlarged graphitic domains surprisingly reveals that the water transport is damped with increasing size of graphitic domains. Supported by AMID simulations and interlayer space measurements we can show that the supressed swelling is the dominant factor for the water flux compared with the increased mobility of water molecules along joined graphitic domains.

Our initial hypothesis that the enhanced graphitic domains should be beneficial for the catalytic activity of GO can be confirmed for original GO. However, in Fe-GO the decline in mass transport upon suppression of swelling plays a more dominant role for the catalytic activity then the improvement through enlarged graphitic domains. By that, we uncover the so far neglected role of mass transport within the GO laminar structure for catalytic activity.

**Materials and methods**

**Preparation of monolayer GO-TEM grids:** 10 µl of highly diluted and sonicated GO solution was dropped on each TEM grid. By that, a sufficient coverage with mono- and few layer GO was accomplished. Distinguishing between mono- and few layers be analysing the diffraction pattern is straight forward and described in detail below. Between heat treatment and analysis, the samples were stored in a desiccator.

**GO films**: GO films were fabricated using the vacuum filtration method as described in our previous studies [32,33]. With a pressure of 60 kPa the GO solution was filtrated through a PVDF support membrane. All the GO films have the same average effective size of ~3.0 cm$^2$. Before further use the films were dried and stored in a desiccator.

**Heat treatment in air and vacuum:** Both, GO-TEM grids and GO films were annealed in vacuum and air but otherwise same conditions. One batch of samples were annealed in a furnace at 80 °C ± 1 °C and covered with a glass container. Another batch of samples were placed in a vacuum oven at 80 °C ± 1° C (see Fig. S10). The temperature was also monitored through multiple spots via a thermocouple inside the oven. The base pressure was 3 x 10$^{-3}$ mbar. To remove residual oxygen and moisture, highly purified N2 was let into the chamber with a mass flow controller so that the pressure is at 1.8 x 10$^{-1}$ mbar. Before initially starting the experiment and after subsequently removing samples from the oven, care was taken that the base pressure was reached and N$_2$ was let into chamber for at least 30 minutes before start heating again.

**Fe-GO for HER and OER:** Aqueous GO solution (0.1 mg/ml) and FeCl$_3$ salt was mixed aiming at a Fe/GO of (1/100). The solution was dried in a vacuum oven at 30 °C and 10 mbar to receive fine Fe-GO powder. The powder Fe-GO were placed in an oven at 80 °C for several days as noted in the sample names.

**Characterization:** HRTEM images were obtained with a Gatan OneView IS camera on a JEOL JEM-F200 which was operated at 80 kV to reduce knock-on damage to the GO. Aberration corrected TEM images and ADF STEM images were obtained at 60 kV on a monochromated FEI

Themis-Z. XPS measurements were carried out on an ULVAC-PHI 5000 Versa probe II, with an Al-Kα monochromatic X-ray source (energy = 1486.68 eV). C 1s = 284.8 eV for adventitious hydrocarbon was used as binding energy reference. XRD patterns were collected with an Empyrean Thin-Film XRD. FTIR spectra were recoded with a PerkinElmer Spectrum 100/Spotlight 400 in attenuated total reflectance. The 13C nuclear magnetic resonance (NMR) experiments were performed using a Bruker AVANCE III 300 spectrometer, with a 7 Tesla superconducting magnet, operating at frequencies of 300 MHz and 75 MHz for the 1H and 13C nuclei, respectively. Approximately 40 mg each of sample was centre packed into 4 mm zirconia rotors fitted with Kel-F ® caps and spun to 12 kHz at the magic angle. The quantitative 13C NMR spectra were acquired with a Hahn-echo sequence to ensure a flat baseline, and 100 s recycle delay to ensure complete signal relaxation. The 1H decoupling was achieved using a SPINAL-64 with a 71 kHz decoupling field strength, and 768-1024 signal transients were co-added to ensure sufficient signal to noise. The 90° pulse lengths of 4 μs and 3.5 μs were used for the 13C and 1H nuclei respectively. The Glycine C=O resonance set to 176.4 ppm was used to reference the NRM chemical shifts.

**Electrochemical experiments**: All electrochemical measurements in this study were carried out with a CHI 760E (CH Instrument, Texas) electrochemical workstation. To prepare electrodes, 5 mg of catalyst powder was dissolved in 0.5 mL deionized water and ethanol solution (1:1, v/v), followed by the addition of 25 μL of Nafion solution (Sigma-Aldrich), and sonicated to form inks, which were then drop-casted on carbon fiber paper to attain a catalyst loading of 0.5 mg cm$^{-2}$. The electrodes were then used as working electrodes, while a Pt wire and saturated calomel electrode was used as the counter and reference electrode, respectively. All potentials measured in this study were converted to the reversible hydrogen electrode (RHE) reference for the purpose of comparison, using the following equation: $E_{RHE}$ (V) = $E_{SCE}$ (V) + 0.245 + 0.059 × pH.

**Water transport measurements** Following the procedure in [16], humidity depend water loss of a container sealed with a GO membrane was recorded. As presented in Fig. S9, saturated salt solutions were placed in containers sealed with GO membranes. Prior to sealing the container, the GO membranes were heat treated for 0, 7 and 14 days. By using saturated salt solutions, the humidity inside the container was controlled to 11 % (LiCl), 21 % ($CH_3COOK$), 55 % ($Mg[NO_3]_2$), 85 % (KCl) and 100 % (pure DI water) RH. By recording the weight loss of these containers, the humidity dependent water flux through the membranes was obtained. For the time of monitoring the weight loss, the containers were placed in a glovebox with humidity <1 % RH and the temperature was constantly monitored and stable at 22.5 ± 0.5 °C. Before starting the experiment, the sealed containers with membranes were kept for 1 week inside the glove box to ensure that the membranes have stabilized.

**Ab initio molecular dynamics (AIMD) simulations** All AIMD calculations were carried out using the plane-wave DFT code, VASP [34,35]. The projector augmented wave (PAW) method is used to treat the core electrons [36], and the exchange-correlation functional was set to Perdew-Burke-Ernzerhof (PBE).[37] The wavefunctions were expanded with a kinetic energy cut-off value of 450 eV and a gamma-point k-grid were used.

We employed GO structures that were utilized in ref. [6], where extensive details on the structures can be found. We used a subset of these GO structures with an epoxy to hydroxyl ratio of 3:2, and an oxygen concentration of 20 at.% in this study. For simulating graphitic domain lateral sizes of 0 and 1.3 nm, we utilized GO structures where oxygen was dispersed uniformly across the graphene plane and where the same oxygen content was concentrated in a smaller domain, as shown in the main text. In these simulations, the interlayer spacing was fixed at 1.5 nm. On the other hand, for simulating different interlayer spacings, we used GO structures with interlayer spacings of 1.0 and 1.5 nm (carbon-carbon plane distance), while the graphitic domain size was fixed (0 nm case). A total of 30 water molecules were inserted within the interlayer space in all our simulations and statistics for determining mean-square displacement (MSD) was extracted on these molecules. MD simulations were carried out using the NVT ensemble using a Nose-Hoover thermostat set to 310 K. A time-step of 1 fs was used, and the simulations were carried out for 2 ps.

**Supporting Information**

Additional Information on TEM study, Supplementary structural and chemical analysis of annealed GO, Additional discussion of the nature of the thermal enhancement of graphitic areas and its benefits for the room temperature stability of GO, Additional discussion and information on study of hydration behaviour of annealed GO, Characterization of Fe-GO

**Acknowledgements**

T.F. acknowledges the UNSW Scientia Ph.D. Scholarship and SSEAU Scholarship. The authors also thank Abdul Hakim and Fei Wang for helpful discussion. R.D. and R.A. acknowledge funding from the Australian Research Council (ARC) Training Centre for Global Hydrogen Economy (IC200100023). R.D., R.A., and R.J. acknowledge funding from Digital Futures Grid Institute at UNSW Sydney. The authors acknowledge the facilities and the scientific and technical assistance of Microscopy Australia at the Electron Microscope Unit (EMU) within the Mark Wainwright Analytical Centre (MWAC) at UNSW Sydney and at The University of Sydney. G.H.L. acknowledge supports from the National Research Foundation (NRF) of Korea (2016M3A7B4910940, 2018M3D1A1058793, 2017R1A5A1014862 (SRC Program: vdWMRC Center)) and Research Institute of Advanced Materials, Institute of Engineering Research, and Institute of Applied Physics.

Supporting Information

# Enhanced graphitic domains of unreduced graphene oxide and the interplay of hydration behaviour and catalytic activity

*Tobias Foller[1], Rahman Daiyan[2], Xiaoheng Jin[1], Joshua Leverett[2], Hangyel Kim[3], Richard Webster[4], Jeaniffer E. Yap[1], Xinyue Wen[1], Aditya Rawal[4], K. Kanishka H. DeSilva[5], Masamichi Yoshimura[5], Heriberto Bustamante[6], Shery L.Y. Chang[1, 4], Priyank Kumar[2], Yi You[1,7], Gwan-Hyoung Lee[3], Rose Amal[2] and Rakesh Joshi[1,] \**

[1]School of Materials Science and Engineering, University of New South Wales, Sydney, NSW 2052, Australia

[2]Particles and Catalysis Research Laboratory and School of Chemical Engineering, University of New South Wales, Sydney, NSW, 2052, Australia

[3]Department of Materials Science and Engineering Research Institute of Advanced Materials (RIAM), Seoul National University, Seoul 08826, Korea

[4] Electron Microscopy Unit, Mark Wainwright Analytical Centre, University of New South Wales, Sydney, NSW, 2052, Australia

[5]Surface Science Laboratory, Toyota Technological Institute, Nagoya 468-8511, Japan

[6]Sydney Water, Parramatta, New South Wales 2125, Australia

[7]School of Physics and Astronomy, University of Manchester, Manchester M13 9PL, UK

*email: r.joshi@unsw.edu.au

# 1. TEM study

## 1.1 Identification of monolayer GO:

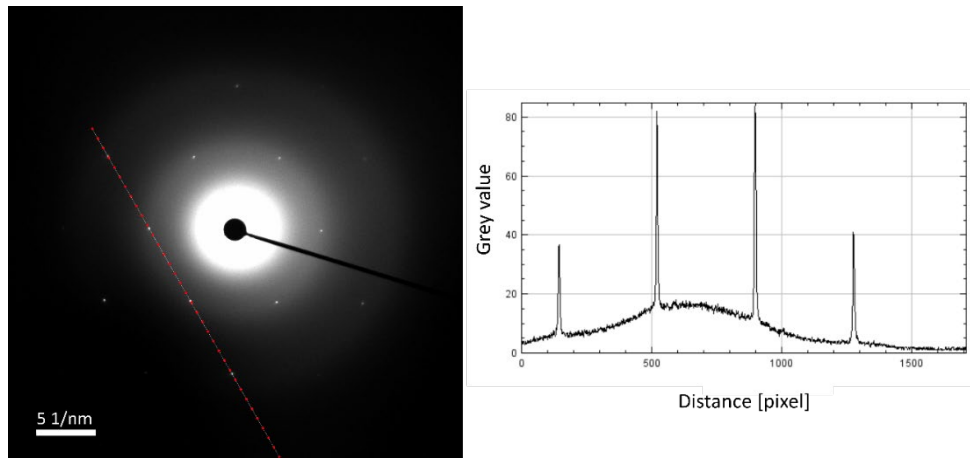

**Figure S1** Typical diffraction pattern of the regions used for analysis. It indicates the presence of monolayer GO. A line profile (red dotted line) shows that the {1-210} type reflections intensity is lower than the {0-110} type reflections intensity. With no significant sample tilt, this suggests the presence of monolayer GO as discussed in previous studies[1,16,17]

## 1.2 Aberration corrected TEM of 14 days heat treated GO in air:

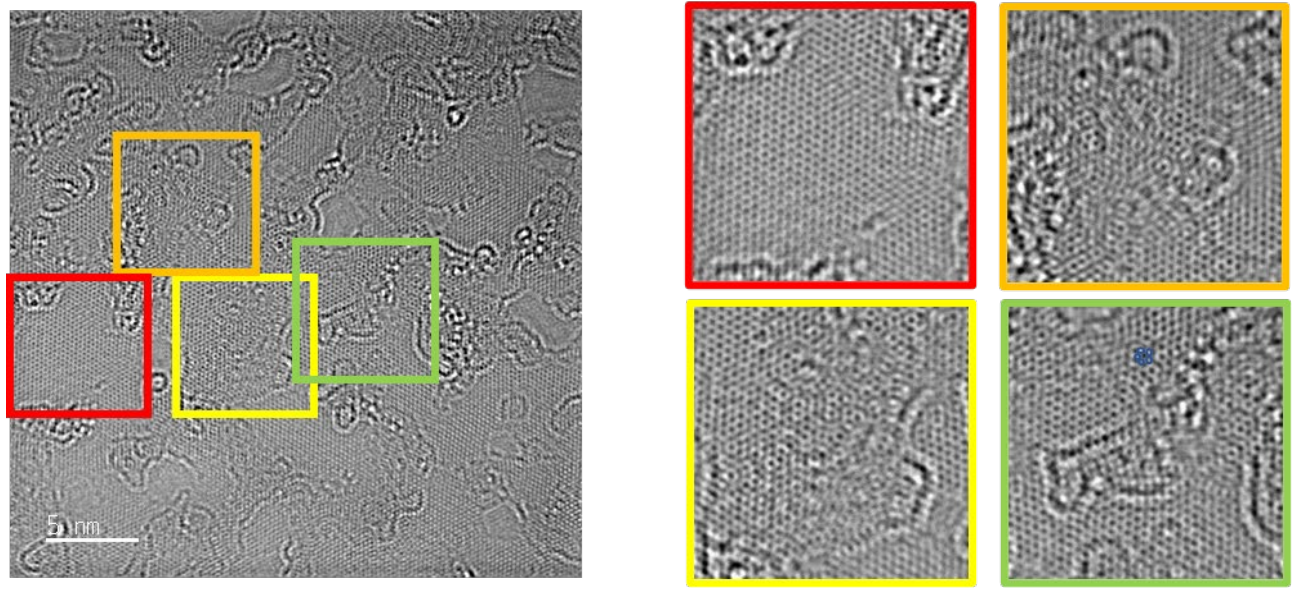

**Figure S2** Aberration corrected TEM picture of GO after 14 days of heat treatment in air. The GO shows large graphitic areas some of which are pristine (red) or show different kinds of mixtures with defective and functionalized areas (yellow, orange, green). Some areas show crystalline defects such as 5-rings (marked with blue rings).

## 1.3 Visual processing of GO TEM images to analyse size and number of graphitic domains for statistical approach

For the histograms in Figure 1C and 1D, as well as each data point shown in Figure S4A and S4B, 6-12 TEM images were acquired and analysed. Supplementary Figure S3 exemplary shows the visual processing of such a TEM image. All images have the same size of 27x27 nm. The graphitic domains are manually marked in light blue. By converting the image into a black and white image, it becomes processable with the particle analysing function of the software Image J, returning a list with size and number of the graphitic domains. By that, the total coverage with graphitic area as well as the size distribution and number of graphitic domains per image are measured and visualized in Fig 1C, 1D and S4A,B.

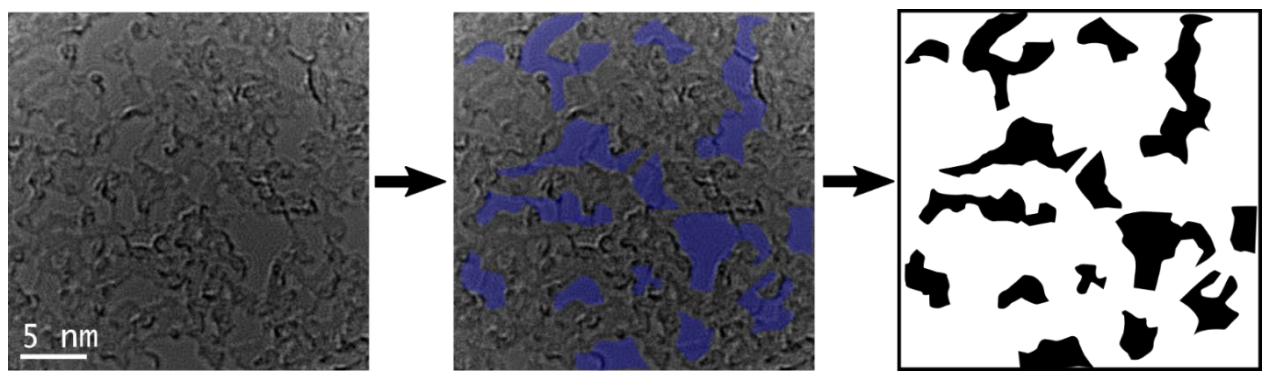

**Figure S3** Visual processing of TEM images to obtain average coverage, size distribution and number of graphitic areas

## 1.4 Development of GO upon electron exposure

*Video S1* shows the development of a GO single layer sheet over the course of 80 s. The images were taken in two second intervals and the video shows 8 frames per second. Care must be taken in the interpretation of such images, since the distinction between oxygen functional groups and other adsorbates, such as hydrocarbons are not directly possible[1–3]. It is apparent that the overall graphitic area does not change significantly over the course of the recorded time. Knock-on damage caused by the electron beam interaction with the sample leads to displacement and desorption of lightly bound adsorbates such as hydrocarbons and physiosorbed oxygen on the graphitic areas. If those adsorbates were dominantly present, the exposure would lead to an increase in graphitic area over the time the images are taken[1,4]. This is not the case. However, movements of the graphitic domain borders and rare creation of holes are observable, matching the argumentation and observation[1] by Erickson et al.[1]. Some adsorbates might still be bound to the oxidized areas, but this does not compromise with any of the analysis or conclusions made in Fig. 1 of the main text since they only rely on the distinction between graphitic domains and functionalized areas.

**1.5 Additional statistical analysis of TEM study:**

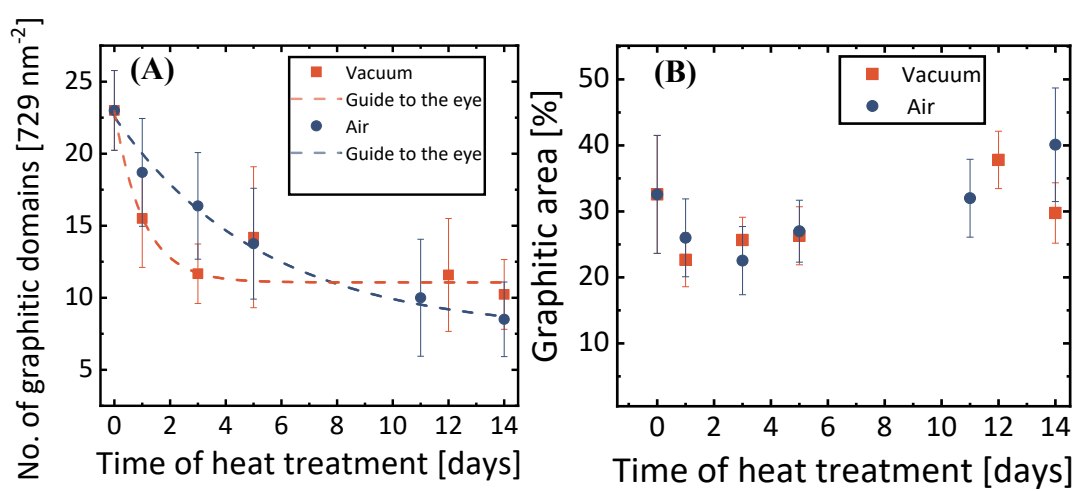

**Figure S4: Statistical analysis of TEM study on heat treated GO.** (a) Exponential decay of the number of separate graphitic domains (b) Average percentage coverage of graphitic area stays constant in vacuum and air.

---

[1] Some studies surprisingly report a strong reduction of GO under repeated imaging[15]. In contrast, our study and work by Erickson et al. does not show such strong changes in GO during the imaging process. It may be attributed to different synthesis methods leading to different composition of the GO. However, this needs further investigation.

## 2. Supplementary structural and chemical analysis

### 2.1 XPS and FTIR

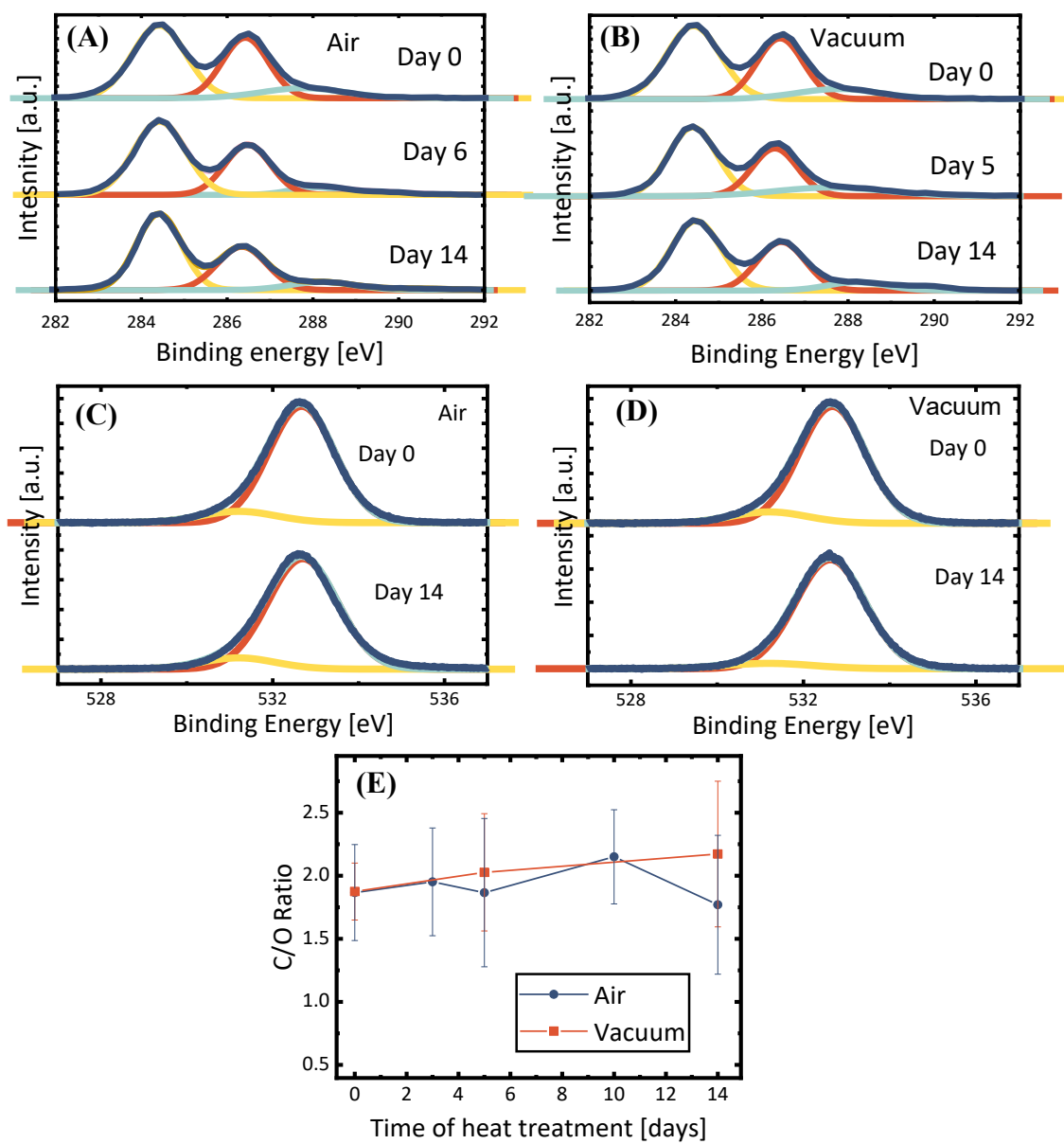

**Figure S5 XPS analysis of GO films heat treated in air and vacuum.** (A/B) C1s peak development for heat treatment in air/vacuum. (C/D) O1s peak development after heat treatment in air/vacuum. (E) C/O ratio time-course for heat treatment in air and vacuum.

Fig. S5 a-e show the XPS spectra over the course of heat treatment. Fig S5A-B show the C1s peak of heat-treated GO in air and vacuum. The results for heat treatment in vacuum are similar. The C-C slightly gains intensity compared to the C-O and C=O during both heat treatment processes. One can see that the shape and position of the O1s peak is mostly unaffected by the heat treatments (see Fig.S5 C and D). In line with that, the C/O ratio in Fig. S5E, which also does not significantly change over the time of heat treatment in air and vacuum. The C/O ratios were calculated as a mean value of at least five different survey scans for each sample. The standard deviation results in the given error bars. The slight increase of the C-C peak may be attributed to an increase in graphitic area and the unchanged O1s peak suggests that the chemically bound oxygen remains qualitatively unchanged upon heat treatment in both environments[5,6].

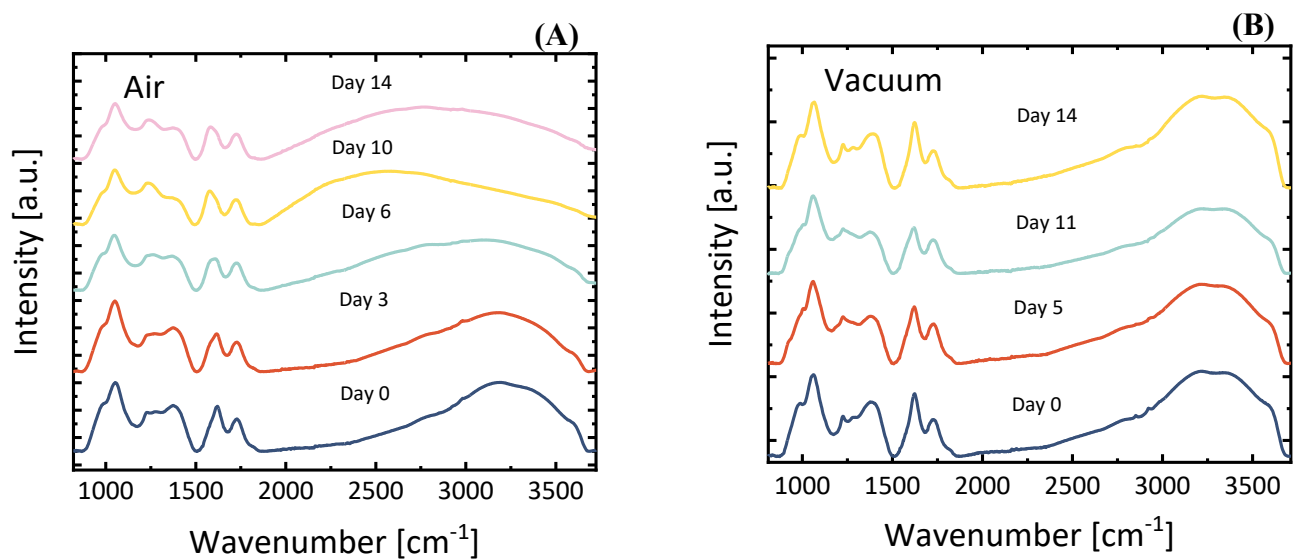

**Figure S6** A/B FTIR spectra for heat treated GO films in air/ vacuum.

Fig. S6 shows the FTIR spectra of heat treated GO in air and vacuum. Generally, the broad peak around 3000-3500 cm$^{-1}$ is attributed to -OH groups. Between 1618 cm$^{-1}$ and 1625 cm$^{-1}$ three peaks overlap from intercalated $H_2O$, C=C and C=O. The region of 1000 cm$^{-1}$ and 1300 cm$^{-1}$ can be assigned to C-O and C-O-C functionals[5–7]. The broad peaks around 3000-3500 cm$^{-1}$ undergo a change and reduction during the 14 days heat treatment period in air, the peak assigned to C=C gains in intensity. In contrast, the vacuum heat-treated samples show no changes in the -OH functional groups. However, a slight change in the composition of the C-O and C-O-C functionals is observable. It may be noted, that in both cases the signature of the functional groups remains largely intact.

**2.2 XRD**

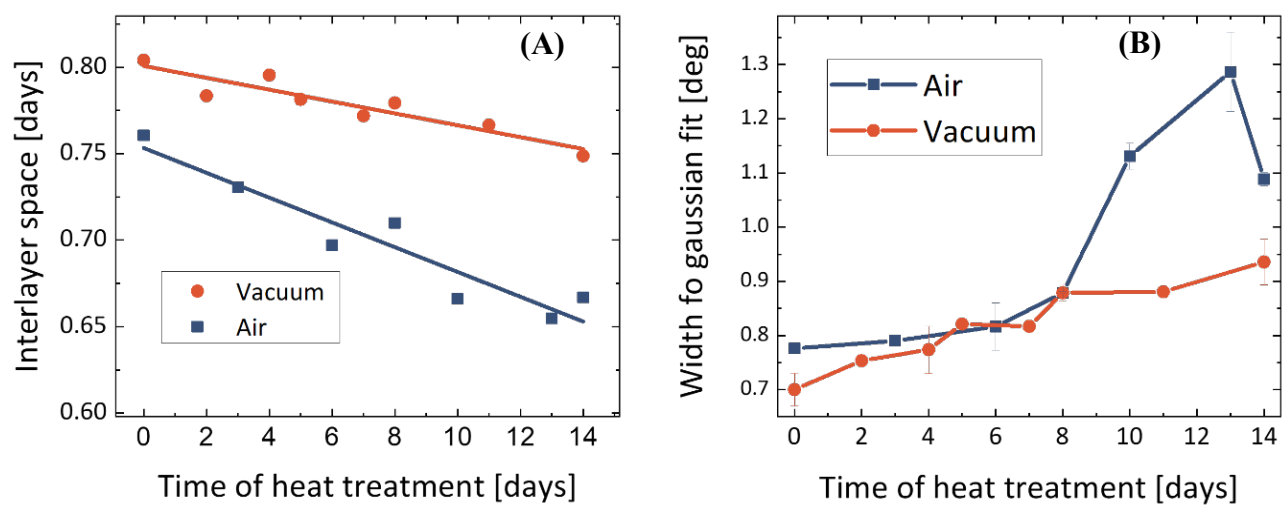

**Figure S7** Analysis of XRD pattern over the course of heat treatment in air and vacuum.

Fig. S7A shows the development of the interlayer space upon heat treatment in vacuum (red) and air (blue). Each dot was measured by fitting the (001) peak of the XRD spectrum recorded for each sample. The angle 2Θ was then converted into an interlayer space by using Bragg's law and the X-ray wavelength of 1.54 Å.

Both heat treatments result in a smaller interlayer space. However, the treatment in air leads to a more pronounced reduction in interlayer space, resembling in a difference in interlayer space of about 0.1 nm between vacuum and air treated samples after 14 days of heat treatment. Moreover, as shown in Fig. S7B, the width of the corresponding gaussian fits increase in air and vacuum with a stronger increase in air treatment.

The observed decrease in interlayer spacing that can also be explained by enlarged graphitic areas (see Fig. 1). Functional groups in GO act as pillars between two adjacent GO layers (29).

Upon increasing their distance by agglomeration, larger graphitic areas begin to sag and decrease the average interlayer space. As a side effect, the range of interlayer spaces becomes wider, resulting in an increase in the width of the XRD peak (Fig. S7B). Previous studies attributed the decrease in interlayer space to the loss of intercalated water. In this case that might be true to a certain degree as well. However, the more pronounced changes in air than in vacuum contradict with this explanation. We want to add that the observed changes in the composition of functional groups might also play a role in changes of interlayer spacing.

**2.3 Colour change upon heat treatment**

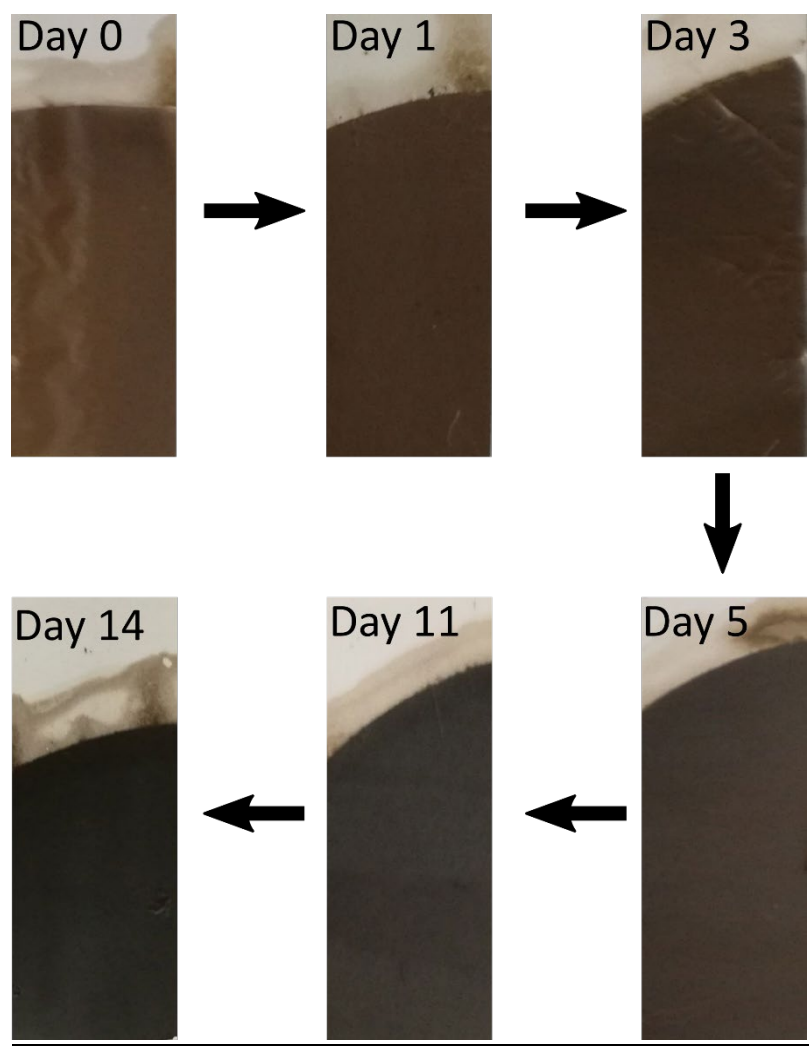

**Figure S8** Photograph series of GO films as days per heat treatment. With increasing time spent at 80 °C, the color of the GO film becomes darker. This indicates a phase transition.

**3. Additional discussion of the nature of the thermal enhancement of graphitic areas and its benefits for the room temperature stability of GO**

A closer look on the SSNMR and TEM results can reveal more details on the nature of the agglomeration process. According to Zhou et al. neighbouring functional groups can form $O_2$ and $H_2O$ molecules, by decomposition reactions of neighbouring functional groups. However, in clusters of functional groups this decomposition is endothermic and further damped by geometrical factors [8]. Thus, the reaction at 80 °C is unlikely and very slow explaining the observed moderate decomposition rate in this study. As it was further predicted by Zhou et al., the most effective low-temperature decomposition is between pairs of epoxy and hydroxyl on the same side of graphene. Consequently, the possibility for those effective pairs to meet would decrease with increasing size of the oxidized areas. Hence, the decomposition probability of functional groups is reduced [8]. The correlation between increasing size in oxidized regions (agglomeration), observed in TEM and decreasing decomposition rate, observed in SSNMR indicates a confirmation of this hypothesis. The fact that the decomposition can be primarily attributed to a decrease in epoxy groups underlines the rational of this explanation.

In more detail, previous theoretical and experimental studies [7–11] suggest that the slight decomposition of functional groups after agglomeration is accompanied by a change in the composition of functional groups in the air atmosphere. In SSNMR, we observe a slight increase in hydroxyl groups by 3 %, while the epoxy groups are decomposed by 7 %. This fits well with predictions based on DFT calculations from Kim et al. that the epoxy groups are more prone to decomposition compared to hydroxyl. Moreover, a transition from hydroxyl to epoxy is energetically favoured[11]. SSNMR studies on $^{13}$C-labelled GO could give more insights into the possible role of edge functionalities in this process[10,12].

# 4. Hydration behaviour of annealed GO

## 4.1 Additional discussion on swelling behaviour

Our combined experimental and computational results allow to explain the hydration behaviour of GO membranes from a new perspective. Despite a constant C/O ratio, substantial changes in the hydration behaviour are observed. In line with a recent study [13] our experiments show a decrease in wet interlayer space and water flux for heat treated membranes. It is straightforward to argue that a smaller interlayer space leads to a decrease in water flux. However, it is not as simple to understand the suppression of swelling. As discussed above, our TEM study shows that the heat treatment leads to enhancement of graphitic domains. According to our simulations enhancement of graphitic domains leads to a higher mobility of water molecules in between the GO nanochannels (Fig. 3c) . The enhanced transport due to higher mobility of water molecules decreases the amount of water that remains inside the channels at a certain time. Consequently, the swelling of GO membranes is reduced. On the other hand, our simulations show that a smaller interlayer space reduces the mobility of water inside the nanochannels. This again would increase the amount of water in the channels at a certain time. These two competing mechanisms govern the size of wet interlayer space and consequently the water flux. Both of which are a direct consequence of the arrangement of functional groups. This may be seen as a tool to tailor GO membranes besides the established methods such as chemical crosslinkers, physical confinement or reduction. For example, using a crosslinker that preserves the interlayer spaces after heat treatment might substantially increase the water flux. This is compelling since most crosslinkers rely on organic bonds which could width stand 80 °C heat treatment. It was suggested that an esterification of adjacent sheets might be responsible for the supressed swelling of heat treated GO [13]. However, this is not in line with our NMR observation showing a slight decrease in ester bonds after heat treatment. It may be noted that an increase in van-der-Waals (vdW) attraction due to the enlarged graphitic domains may also contribute to the suppression of swelling [14]. Further research assessing the changes of refractive index and dielectric constant upon heat treatment of GO to estimate the vdW attraction with a simplified parallel surface model using the Hamaker constant might give more insight.

**4.2 Water flux measurements**

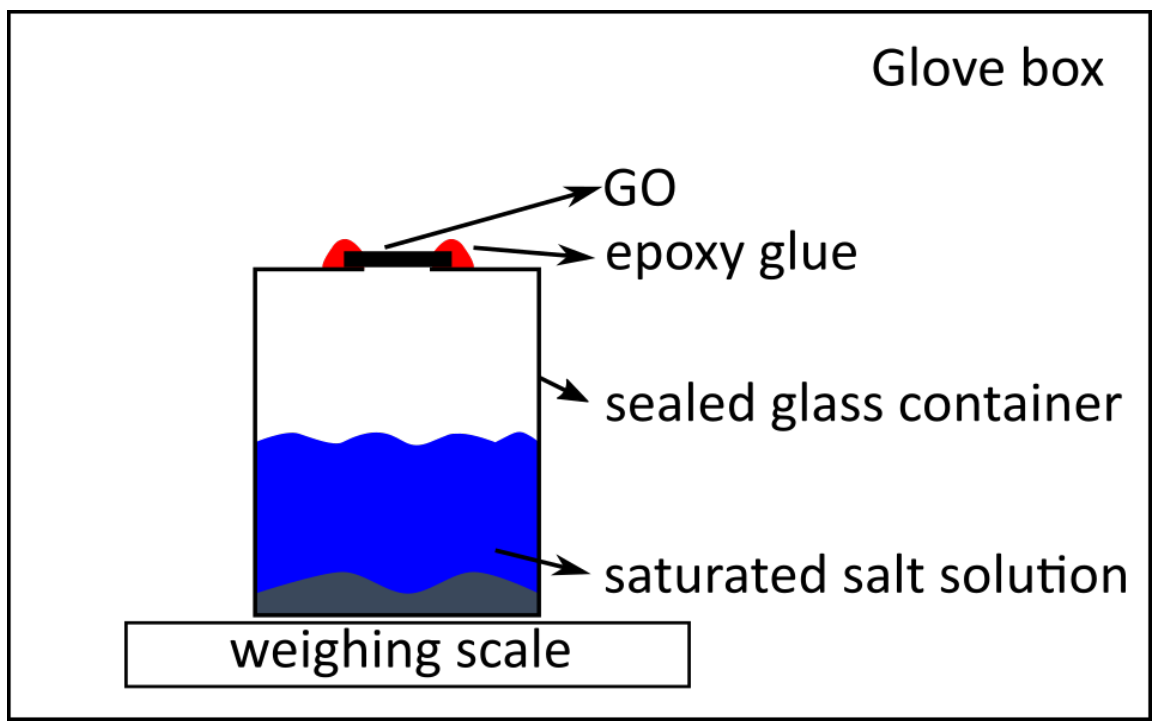

**Figure S9 Schematic of the expermiental set-up for measuring the water transport through original and heat treated GO membranes.** The membranes were used to seal a container with saturated salt solutions. By that, the humidity dependend water loss can be recorded by measuring the weight loss over time. To ensure stable experimental conditions and that the GO membranes are only exposed to the controled humidity of the salt solution, the experiments were carried out in a glove box with dry $N_2$ enviroment (<0.5 % RH).

**5 Characterization of Fe-GO**

**5.1 Stability of Fe-intercalation in Water**

A Fe-GO film on PVDF was created by vacuum filtration of 20 ml Fe-GO solution. The membrane was immersed into 5 ml of DI water for 7 days. The DI water was then checked for Fe ion concentration. The detected concentration is neglectable and below the detection limit (0.05 mg/l). In other words, less then 0.25 μg of Fe ions left the membrane upon immersion in water for 7 days. By filtration of 20 ml Fe-GO solution, 2 mg of Fe-GO was placed onto the PVDF membrane. Keeping in mind that the Fe-GO solution has ~1 wt% Fe ions inside, the sample contained approx. 20 µg of Fe ions. Consequently > 98.75 % of Fe ions stayed inside the membrane over the course of 7-day immersion in DI-water.

**5.2 ADF_STEM characterization of Fe-GO**

Aberration-corrected STEM (Themis, Thermofisher) operated at 60 kV was used to acquire the ADF-STEM images of Fe-GO. The TEM specimens were prepared following the procedure described in the previous session. The imaging was carried out using aberration-corrected annular dark field (ADF) STEM mode where the probe size is approximately 0.14 nm. The dwell time of the scanning beam was kept deliberately low and high spot size was used to reduce the total electron dose on the specimen. This condition allows sufficient spatial resolution for imaging single Fe atom, and the contrast mechanism is sensitive to the high atomic (Z) number element, therefore allowing differentiating the locations of single Fe atoms.

**6. Sample preparation**

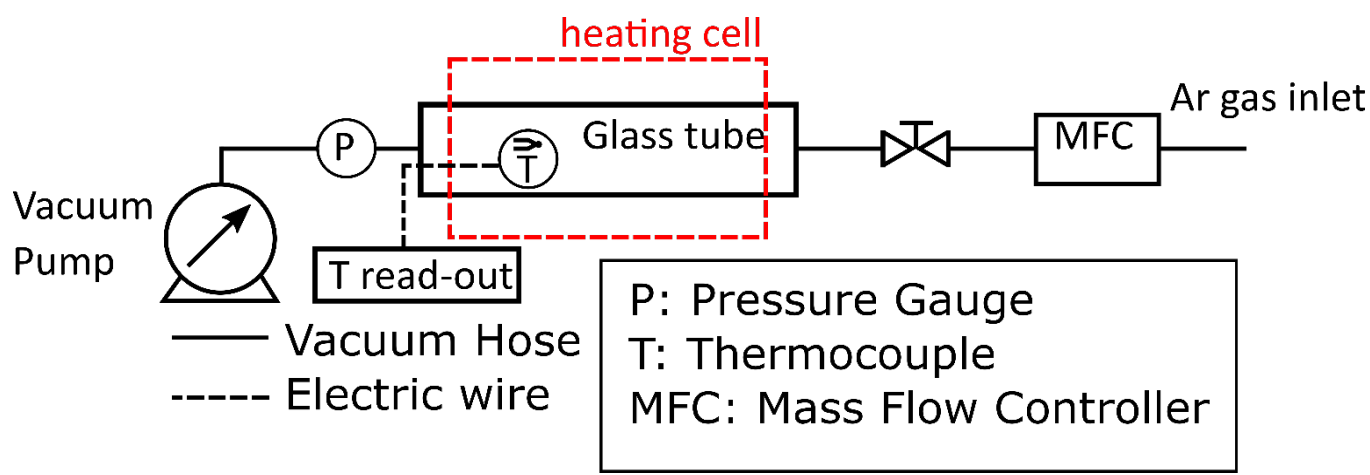

*Figure S10: Experimental Set-up for vacuum heat treatment of GO samples.*